\documentclass[aps,prd,twocolumn,showpacs,superscriptaddress,nofootinbib]{revtex4}
\usepackage{aas_macros}
\usepackage{amsmath}
\usepackage{amssymb}
\usepackage{graphicx}
\usepackage{hyperref}
\usepackage{physics}
\usepackage{axodraw2}
\usepackage{diagrules}
\usepackage[capitalize]{cleveref}

\hypersetup{
    bookmarks=true,         
    unicode=false,          
    pdftoolbar=true,        
    pdfmenubar=true,        
    pdffitwindow=false,     
    pdfstartview={FitH},    
    pdfnewwindow=true,      
    linkcolor=red,          
    citecolor=cyan,        
    filecolor=magenta,      
    urlcolor=blue,           
    linktocpage=true
}

\newlength{\figw}
\newlength{\figh}
\newcommand{\heaviside}[1]{\mathrm{\Theta}\!\left( #1 \right)}

\newcommand{\hermite}[2]{H_{#1}\!\left(#2\right)}
\newcommand{\hermiteb}[2]{H_{#1}\!\left[#2\right]}

\newcommand{\uK}{\mathrm{K}}

\newcommand{\us}{\mathrm{s}}
\newcommand{\uinf}{\mathrm{inf}}

\newcommand{\uT}{\mathrm{T}}

\newcommand{\calP}{\mathcal{P}}

\newcommand{\OmegaK}{\Omega_{\uK}}
\newcommand{\OmegaKbar}{\bar{\Omega}_{\uK}}

\newcommand{\OmegaKzero}{\Omega_{\uK_\zero}}

\newcommand{\bx}{\vb*{x}}
\newcommand{\bk}{\vb*{k}}
\newcommand{\bp}{\vb*{p}}
\newcommand{\bq}{\vb*{q}}
\newcommand{\bSigma}{\vb*{\Sigma}}
\newcommand{\bXi}{\vb*{\Xi}}

\newcommand{\Sigmaw}{\Sigma_\omega}
\newcommand{\omegakbar}{\bar{\omega}}

\newcommand{\atilde}{\tilde{a}}
\newcommand{\Htilde}{\tilde{H}}

\newcommand{\mutilde}{\tilde{\mu}}

\newcommand{\zetac}{\xi}
\newcommand{\zetas}{\zeta_{\us}}
\newcommand{\ksigma}{k_{\sigma}}
\newcommand{\keps}{k_{\varepsilon}}

\newcommand{\zero}{{0}}
\newcommand{\tauzero}{\tau_\zero}
\newcommand{\Hzero}{H_{\zero}}
\newcommand{\azero}{a_{\zero}}

\newcommand{\calPzeta}{\calP_\zeta}
\newcommand{\Pzeta}{P_\zeta}
\newcommand{\calPstar}{\calP_*}
\newcommand{\Ninf}{N_\uinf}

\newcommand{\efolds}{$e$-folds}

\begin{document}

\title{Spatial Curvature from Super-Hubble Cosmological Fluctuations}

\author{Baptiste Blachier}
\email{baptiste.blachier@ens-lyon.fr}
\affiliation{Cosmology, Universe and Relativity at Louvain (CURL),
  Institute of Mathematics and Physics, University of Louvain, 2 Chemin
  du Cyclotron, 1348 Louvain-la-Neuve, Belgium}
\affiliation{Department of Physics, \'Ecole Normale Sup\'erieure de Lyon,
  69364 Lyon, France}

\author{Pierre Auclair}
\email{pierre.auclair@uclouvain.be}
\affiliation{Cosmology, Universe and Relativity at Louvain (CURL),
  Institute of Mathematics and Physics, University of Louvain, 2 Chemin
  du Cyclotron, 1348 Louvain-la-Neuve, Belgium}

\author{Christophe Ringeval}
\email{christophe.ringeval@uclouvain.be}
\affiliation{Cosmology, Universe and Relativity at Louvain (CURL),
  Institute of Mathematics and Physics, University of Louvain, 2 Chemin
  du Cyclotron, 1348 Louvain-la-Neuve, Belgium}

\author{Vincent Vennin}
\email{vincent.vennin@ens.fr}
\affiliation{Laboratoire de Physique de
  l'\'Ecole Normale Sup\'erieure, ENS, CNRS, Universit\'e PSL,
  Sorbonne Universit\'e, Universit\'e Paris Cit\'e, 75005 Paris,
  France}

\date{\today}

\begin{abstract}
We revisit how super-Hubble cosmological fluctuations induce, at any
time in the cosmic history, a non-vanishing spatial curvature of the
local background metric. The random nature of these fluctuations
promotes the curvature density parameter to a stochastic quantity for
which we derive novel non-perturbative expressions for its mean,
variance, higher moments and full probability distribution. For
scale-invariant Gaussian perturbations, such as those favored by
cosmological observations, we find that the most probable value for the
curvature density parameter $\OmegaK$ today is $-10^{-9}$, and that its
mean is $+10^{-9}$, both being overwhelmed by a standard deviation of
order of $10^{-5}$. We then discuss how these numbers would be affected
by the presence of large super-Hubble non-Gaussianities, or if
inflation lasted for a very long time. In particular, we find that
substantial values of $\OmegaK$ are obtained if inflation lasts for
more than a billion {\efolds}.
\end{abstract}

\maketitle

\section{Introduction}

Cosmic structures in the Universe are understood to be seeded by some
pre-existing super-Hubble cosmological fluctuations. Their
gravitational collapse starts when their size becomes smaller than the
Hubble radius, an inevitable outcome in any decelerating
Friedmann-Lema\^itre spacetime. Observational evidence of this
mechanism is present in the Cosmic Microwave Background (CMB) data by
the correlation patterns associated with the polarization and
temperature angular power spectra~\cite{Hu:1996yt, Planck:2018nkj}, as
well as in the statistics of the large-scale structures observed at
lower redshifts~\cite{Fonseca:2015laa, EUCLID:2020jwq}.

Cosmic Inflation, an early era of accelerated cosmic expansion, is the
prime candidate to explain the origin of the super-Hubble
fluctuations. They are of quantum origin, stretched to length scales
much larger than the Hubble radius during
inflation~\cite{Starobinsky:1979ty, Starobinsky:1980te, Guth:1980zm,
  Linde:1981mu, Albrecht:1982wi, Linde:1983gd, Mukhanov:1981xt,
  Mukhanov:1982nu, Starobinsky:1982ee, Guth:1982ec, Hawking:1982cz,
  Bardeen:1983qw}. At the same time, inflation smooths out any
pre-existing inhomogeneity, and one of the historical motivations for
Cosmic Inflation is that the spatial curvature of spacetime,
$\OmegaK$, should be exponentially small at the end of inflation (at
most $e^{-60}$). This prediction is compatible with the current bound
$|\OmegaKzero|<3 \times 10^{-3}$ today, coming from the Planck CMB
data and Baryon Acoustic Oscillations (BAO) measurements.

Intuitively, the existence, today, of Hubble-sized curvature
fluctuations suggests that these could be confused with a small
non-vanishing spatial curvature of the local background metric. In
particular, these modes are expected to induce a limitation on our
ability to measure very small values of the curvature density
parameter~\cite{Waterhouse:2008vb, Buchert:2008zm, Vardanyan:2009ft,
  Leonard:2016evk, Anselmi:2022uvj}. More than being a nuisance, we
will show that super-Hubble (hence, ``conserved'') fluctuations do
create spatial curvature.

In order to deal with fluctuations over a background metric when both are
intertwined, we can start from the inhomogeneous metric proposed in
Refs.~\cite{Salopek:1990jq, Creminelli:2004yq, Kolb:2004jg, Lyth:2004gb}:
\begin{equation}
  \dd s^2 = -\dd \tau^2 + a^2(\tau) e^{2 \zeta(\tau,\bx)}
    \delta_{ij}\dd x^i \dd x^j.
\label{eq:metric}
\end{equation}
This metric is not fully general, as inhomogeneities are all contained
in one scalar function $\zeta$. However, as discussed in
Refs.~\cite{Salopek:1990jq, Creminelli:2004yq, Kolb:2004jg,
  Lyth:2004gb}, this is the most generic metric in the absence of vector-
and tensor-type inhomogeneities, and in the gauge where fixed time
slices have uniform energy density and fixed spatial worldlines are
comoving with matter. At super-Hubble scales, this reduces to
the synchronous gauge supplemented by some additional conditions that
fix it uniquely. The quantity $\zeta(\tau,\bx)$ can be shown to
be ``conserved'' at large distances. As such, it provides a non-linear
generalization of the constant-energy-density curvature
perturbation~\cite{Rigopoulos:2004gr, Langlois:2005ii}.

Historically, this metric has been intensively discussed in the
attempts to explain the acceleration of the Universe by the
backreaction of super-Hubble
inhomogeneities~\cite{Kolb:2005me,Barausse:2005nf}. But, as realized
soon after~\cite{Hirata:2005ei, Kolb:2005da, Geshnizjani:2005ce}, the
effects of super-Hubble fluctuations onto the background evolution are
to modify the spatial curvature. Let us notice that, on top of the
background evolution, other observable signatures are
possible~\cite{1978SvA....22..125G, Garcia-Bellido:1995fgh,
  Erickcek:2008jp}. To our knowledge, the only works having
addressed how super-Hubble modes affect the spatial curvature are
Refs.~\cite{Brandenberger:2004ix, Geshnizjani:2005ce, Kleban:2012ph},
based, however, on perturbative gradient expansions or linear
perturbation theory only. When the non-perturbative terms of our
derivation can be neglected, we recover some of their results.

The paper is organized as follows. In \cref{sec:curvdef}, we derive an
exact expression for the curvature density parameter $\OmegaK$ in
terms of the non-linear curvature perturbation $\zeta$. This promotes
$\OmegaK$ to a stochastic quantity, and in \cref{sec:stats} we
calculate its moments as well as its probability density function,
assuming Gaussian statistics for $\zeta$. Finally, we conclude by
discussing how the statistics of the curvature density parameter is
modified in the presence of non-Gaussian super-Hubble fluctuations or
if inflation lasted for a very long time.

\section{Curvature density parameter}
\label{sec:curvdef}

When spatial curvature is included, the Friedmann-Lema\^{\i}tre-Robertson-Walker
(FLRW) line element reads
\begin{equation}
  \dd s^2 = -\dd \tau^2 + a^2(\tau) 
  \frac{  \delta_{ij}\dd x^i \dd x^j}{\left(1+\frac{K}{4}\delta_{mn}x^m x^n\right)^2}\,,
\label{eq:metric:curved:FLRW}
\end{equation}
where $K$ is a constant, and its Ricci scalar is given by
\begin{equation}
\label{eq:Ricci:curved:FLRW}
R=6\frac{\dot{a}^2}{a^2}+6\frac{\ddot{a}}{a}+\frac{6}{a^2}K .
\end{equation}

The metric \eqref{eq:metric} can be viewed as an inhomogeneous
generalization of a flat, i.e., $K=0$, FLRW spacetime having a
space-dependent scale factor
\begin{equation}
b(\tau,\bx) \equiv a(\tau) e^{\zeta(\tau,\bx)},
\end{equation}
from which one can derive the Ricci scalar
\begin{equation}
R = 6 \dfrac{\dot{b}^2}{b^2} + 6
\dfrac{\ddot{b}}{b} + 2 \dfrac{(\vnabla
  b)^2}{b^4} - 4 \dfrac{\Delta b}{b^3}\,.
\label{eq:ricci}
\end{equation}
We now split $\zeta(\tau,\bx)=\zetac(\bx) + \zetas(\tau,\bx)$ into a
conserved part $\zetac$ (super-Hubble) and time-dependent
fluctuations $\zetas$ (sub-Hubble). Expanding in the (presumably
small) short-length part, one has
\begin{equation}
b(\tau,\bx) = a(\tau) e^{\zetac(\bx)} \left[1 + \zetas(\tau,\bx) +
\cdots \right],
\label{eq:pertexpand}
\end{equation}
and upon defining
\begin{equation}
\atilde(\tau,\bx) = a(\tau) e^{\zetac(\bx)}
\end{equation}
one is led to
\begin{equation}
R = 6 \dfrac{\dot{\atilde}^2}{\atilde^2} +
6\dfrac{\ddot{\atilde}}{\atilde} +
\dfrac{6}{\atilde^2}\left[-\dfrac{2}{3} \Delta \zetac -
    \dfrac{1}{3} \left(\vnabla \zetac\right)^2 \right] + \cdots.
\label{eq:ricciexpand}
\end{equation}
The omitted terms in this expression are the ones appearing in the
linear theory of cosmological perturbations, in the synchronous gauge,
completed by all possible non-linear corrections involving powers of
$\zetas(\tau,\bx)$ and products with
$\atilde(\tau,\bx)$~\cite{Carrilho:2015cma}. The mixed terms involving
both $\atilde(\tau,\bx)$ and powers of $\zetas(\tau,\bx)$ were
precisely the ones discussed in the early works on backreaction and
are non-observable~\cite{Geshnizjani:2005ce, Hirata:2005ei,
  Kolb:2005da}. As can be checked in \cref{eq:ricciexpand}, the terms
we have kept are invariant by a constant shift of $\zetac(\bx)$, up to
a redefinition of $a(\tau)$.

Since $\zetac(\bx)$ varies on super-Hubble length scales only, so does
$\atilde(\tau,\bx)$; hence, any observer will identify
$\atilde(\tau,\bx)$ as the FLRW scale factor of their local Hubble
patch. Let us notice that, in the gauge we work in, the Hubble radius
is the same for all observers, since~\cite{Geshnizjani:2002wp,
  Matarrese:2003ye,Kolb:2004jg}
\begin{equation}
\Htilde \equiv \dfrac{\dot{\atilde}}{\atilde} = \dfrac{\dot{a}}{a} =
H,
\end{equation}
which does not depend on $\bx$. An important remark is that
\cref{eq:Ricci:curved:FLRW,eq:ricciexpand} coincide upon identifying
\begin{equation}
K = -\dfrac{2}{3} \Delta \zetac - \dfrac{1}{3} \left(\vnabla \zetac \right)^2,
\label{eq:K}
\end{equation}
which is indeed constant, since $\zetac$ is conserved, 
and whose measurable curvature density parameter reads
\begin{equation}
\OmegaK = -\dfrac{K}{\atilde^2 \Htilde^2} = -\dfrac{K e^{-2\zetac}}{a^2 H^2}\,.
\label{eq:OmegaK}
\end{equation}
Let us stress that Eq.~\eqref{eq:K} is exact in the sense that all the
terms omitted involve $\zetas(\tau,\bx)$; hence, they are time-dependent
and cannot be absorbed in $K$.
Equation~\eqref{eq:K} makes also explicit that only gradients
of super-Hubble inhomogeneities have a non-trivial effect.

\section{Statistics}
\label{sec:stats}

Current cosmological measurements~\cite{Planck:2019kim} imply that
$\zeta$ has Gaussian statistics and can, thus, be treated as a random
Gaussian field, with vanishing mean and higher-point correlation
functions entirely determined by the power spectrum
\begin{equation}
\ev{\zeta(\bk) \zeta(\bk')} = \left(2 \pi\right)^3 \delta(\bk+\bk')\Pzeta(k).
\label{eq:Pk}
\end{equation}
This is also in agreement with the most favored inflationary scenarios, where the
mean values are identified with vacuum expectation values of quantum operators in
the Bunch-Davis vacuum.
Later on, we will also use the spherical power spectrum $\calPzeta(k)$
defined by
\begin{equation}
\calPzeta(k) = \dfrac{k^3}{2\pi^2} \Pzeta(k) \simeq \calPstar,
\end{equation}
where the last approximation holds for a scale-invariant power
spectrum. 

From \cref{eq:K,eq:OmegaK}, $\OmegaK$ can, therefore, also be seen as a
stochastic quantity, though its non-linear dependence on $\zetac$, and,
thus, on $\zeta$, implies that it does not feature Gaussian
statistics. In particular, its expectation value does not necessarily
vanish.

Let us make the decomposition $\zeta(\tau,\bx)=\zetac(\bx) + \zetas(\tau,\bx)$ explicit in
Fourier space:
\begin{equation}
\begin{aligned}
  \zeta(\tau,\bx)  & = \dfrac{1}{\left(2\pi \right)^3} \int\dd^3 \bk \,
  \heaviside{\ksigma-k} \zeta(\bk) e^{i\bk\vdot\bx} \\ & +
  \dfrac{1}{\left(2 \pi \right)^3} \int \dd^3 \bk \heaviside{k-\ksigma}
    \zeta(\tau,\bk) e^{i \bk \vdot \bx},
  \end{aligned}
\label{eq:fouriersplit}
\end{equation}
where we have introduced a wave number $\ksigma$ below which all
Fourier modes $\zeta(\tau,k<\ksigma) = \zeta(k)$ can be approximated
as time independent. Based on the theory of cosmological
perturbations, and its generalizations~\cite{Rigopoulos:2004gr,
  Langlois:2005ii}, this wave number is at most of the order of the
conformal Hubble parameter at the observer's time, say, $\tauzero$;
namely, $\ksigma \lesssim \atilde(\tauzero) \Htilde(\tauzero)$. Let us
remark the presence of $\atilde(\tauzero,\bx)$, instead of
$a(\tauzero)$, in this expression. A \emph{priori}, this would induce an
extra dependence on $\bx$ in \cref{eq:fouriersplit}, where one should
write $\ksigma(\bx)$. In order to circumvent this issue, we can, for
now, simply choose the cutoff $\ksigma$ to be sufficiently small such
that it encompasses all possible spatial modulations of
$\atilde(\tauzero,\bx)$. In other words, we define
\begin{equation}
\ksigma \equiv \sigma \azero \Hzero\,,
\label{eq:ksigma}
\end{equation}
where, in principle, $\sigma<e^{{\min}_{\bx}(\zetac)}$. As such, we can
identify the conserved quantity with
\begin{equation}
\zetac(\bx)=\dfrac{1}{\left(2\pi\right)^3} \int \dd^3\bk \,
\heaviside{\ksigma-k} \zeta(\bk)
e^{i \bk \vdot \bx}\,.
\label{eq:zetacdef}
\end{equation}
Let us remark that $\sigma$ also quantifies the possible ambiguities
in separating the background, made of the time-independent $\zetac(\bx)$, from
the modes which contribute to the perturbations, the time-dependent
$\zetas(\tau,\bx)$.

\subsection{Mean value}

The mean value of the curvature density parameter reads
\begin{equation}
\ev{\OmegaK} = -\dfrac{\ev{K e^{-2 \zetac}}}{a^2 H^2}\,,
\label{eq:vevdefOmegaK}
\end{equation}
where $\zetac$ is given by \cref{eq:zetacdef}.
The curvature scalar $K$, given in \cref{eq:K}, can be split into
two terms $K = K_1 + K_2$ with
\begin{equation}
K_1 \equiv -\dfrac{2}{3} \Delta \zetac, \qquad K_2 \equiv
-\dfrac{1}{3} \left(\vnabla \zetac \right)^2.
\label{eq:K12def}
\end{equation}
Therefore, one needs the Laplacian and the squared gradient 
of $\zetac$. They read, respectively,
\begin{equation}
\Delta \zetac = - \int \dfrac{\dd^3 \bk}{\left(2\pi\right)^3}
\heaviside{\ksigma-k} k^2 \zeta(\bk) e^{i \bk \vdot \bx},
\label{eq:laplacian}
\end{equation}
and
\begin{equation}
  \begin{aligned}
    \left(\vnabla \zetac \right)^2 & = -\int \dfrac{\dd^3
      \bp \dd^3\bq}{\left(2\pi\right)^6} \, \heaviside{\ksigma-p}
    \heaviside{\ksigma-q} \\ & \times  \bp \vdot
    \bq \, \zeta(\bp) \zeta(\bq)
    e^{i\left(\bp + \bq\right)\vdot\bx},
  \end{aligned}
  \label{eq:gradmag2}
\end{equation}
from which one can immediately calculate
\begin{equation}
\begin{aligned}
  \ev{K} & = \ev{K_2} = -\dfrac{1}{3} \int \dfrac{\dd^3
    \bk}{\left(2\pi\right)^3} \heaviside{\ksigma-k} k^2 \Pzeta(k) \\
  & = -\dfrac{1}{3} \int_0^{\ksigma} \dd k k \calPzeta(k) \simeq
  -\dfrac{1}{6} \ksigma^2 \calPstar,
\end{aligned}
\label{eq:evK}
\end{equation}
the rightmost equality holding only for a scale-invariant power spectrum.    

The term $e^{-2\zetac}$ appearing in \cref{eq:vevdefOmegaK} can be
expressed in terms of $\zeta(\bk)$ by using the series representation
\begin{equation}
e^{-2\zetac} = \sum_{n=0}^{+\infty} \dfrac{(-2)^n}{n!} \zetac^n,
\label{eq:expm2zetac}
\end{equation}
with
\begin{equation}
\zetac^n = \int \dfrac{\dd^3 \bk_1 \dots \dd^3 \bk_n}{\left(2
  \pi\right)^{3n}} \left[\prod_{j=1}^n \heaviside{\ksigma - k_j}
\zeta(\bk_j) \right] e^{i \bx \vdot \sum_j \bk_j}.
\label{eq:zetacn}
\end{equation}
As can be seen in \cref{eq:vevdefOmegaK}, the mean value of the
curvature density parameter requires the explicit determination of an
infinite number of terms, the non-vanishing ones being of the form
$\ev{K_1 \zetac^{2p + 1}}$ and $\ev{K_2 \zetac^{2p}}$. From
\cref{eq:Pk,eq:K12def,eq:zetacn}, one can make extensive use of the
Wick theorem to reduce all the expectation values to a few
two-point functions with the following diagrammatic rules:
\begin{equation}
  \begin{aligned}
\ev{K_1 K_1} & \equiv  \diagKoneKone  = \ev{K_1^2},\\
\ev{\zetac \zetac} & \equiv  \diagZetaZeta  = \ev{\zetac^2},\\
\ev{K_1\zetac} & \equiv  \diagKoneZeta  = - 2 \ev{K},\\
 \ev{K_2} & \equiv \diagKtwoSelf  = \ev{K}.
\end{aligned}
\label{eq:diagrules}
\end{equation}
Let us notice that, due to the inner product structure of
\cref{eq:gradmag2}, the $K_2$ vertices have two ``legs'' that can
connect only to other $K_2$ vertices.  From \cref{eq:laplacian}, one has
\begin{equation}
  \ev{K_1^2}  = \dfrac{4}{9} \int_0^{\ksigma} \dd k k^3 \calPzeta(k) \simeq
  \dfrac{1}{9} \ksigma^4 \calPstar,
\label{eq:m2K1}
\end{equation}
which allows us to express the second moment of the curvature scalar
as
\begin{equation}
\ev{K^2} = \ev{K_1^2} + \dfrac{5}{3} \ev{K}^2 \simeq \dfrac{1}{9}
\ksigma^4 \calPstar \left(1+ \dfrac{5}{12} \calPstar\right).
\label{eq:m2K}
\end{equation}
In \cref{eq:diagrules}, we also need the variance of the conserved
quantity $\zetac$. It can be determined from \cref{eq:zetacdef} and reads
\begin{equation}
\ev{\zetac^2} = \int_{\keps}^{\ksigma}\dd k \dfrac{\calPzeta(k)}{k} \simeq
\calPstar \ln\left(\dfrac{\ksigma}{\keps} \right) \simeq \calPstar \Ninf,
\label{eq:varzetac}
\end{equation}
where we have introduced an expected infrared cutoff $\keps$. Indeed,
in the context of Cosmic Inflation, the ratio between the largest and
shortest lengths being amplified is precisely given by the total amount of
stretching generated by the accelerated expansion, the so-called total
number of {\efolds} $\Ninf$. For the measured value of $\calPstar =
2.1\times 10^{-9}$~\cite{Planck:2018jri}, and a not too long
inflationary era $\Ninf \ll 10^{9}$, $\ev{\zetac^2}$ is a small
quantity.

Denoting by $W_{2p}=(2p)!/(p!2^p)$ the number of Wick contractions
between $p$ pairs, one obtains
\begin{equation}
\begin{aligned}
  \ev{K_1 \zetac^{2p+1}} & = (2p+1) \left(\diagKoneZeta\right) \times
  W_{2p} \left(\diagZetaZeta\right)^{p}\\
  & =  -\dfrac{(2p+1)!}{p! \, 2^{p-1}} \ev{K}
  \ev{\zetac^2}^p,
\end{aligned}
\label{eq:evK1zetap}
\end{equation}
and
\begin{equation}
\begin{aligned}
  \ev{K_2 \zetac^{2p}} & = \diagKtwoSelf \times W_{2p}
  \left(\diagZetaZeta\right)^{p} \\
  &= \dfrac{(2p)!}{p! \, 2^p} \ev{K} \ev{\zetac^2}^p.
\label{eq:evK2zetap}
\end{aligned}
\end{equation}
The infinite series obtained by combining
\cref{eq:evK1zetap,eq:evK2zetap,eq:expm2zetac} can be resummed and one gets the
exact expression
\begin{equation}
\ev{\OmegaK} = - \dfrac{5}{a^2 H^2} \ev{K} e^{2 \ev{\zetac^2}}.
\label{eq:evOmegaK}
\end{equation}
Making use of \cref{eq:evK,eq:varzetac}, for a scale-invariant power
spectrum, \cref{eq:evOmegaK} simplifies to
\begin{equation}
\ev{\OmegaKzero} \simeq \dfrac{5}{6} \dfrac{\ksigma^2}{\azero^2 \Hzero^2} \calPstar
e^{2\calPstar \Ninf} \simeq \dfrac{5}{6} \sigma^2 \calPstar,
\label{eq:evOmegaKnum}
\end{equation}
which saturates for $\sigma=1$ at $\ev{\OmegaKzero} \simeq 1.7 \times
10^{-9}$, a barely open universe were we to interpret this number within a
FLRW metric with trivial topology.

\subsection{Variance}

There is little hope to measure such a small value of $\ev{\OmegaK}$,
but, $\OmegaK$ being a stochastic variable, its realizations are also
dictated by the higher moments, the second one being given by
\begin{equation}
\ev{\OmegaK^2} = \dfrac{\ev{K^2 e^{-4 \zetac}}}{a^4 H^4} =
\dfrac{\ev{\left(K_1^2 + 2 K_1 K_2 + K_2^2\right)e^{-4 \zetac}}}{a^4 H^4}\,.
\label{eq:m2defOmegaK}
\end{equation}
Using again a series representation for the exponential,
\cref{eq:m2defOmegaK} can be expanded in an infinite sum requiring the
calculation of the non-vanishing terms $\ev{K_1^2 \zetac^{2p+2}}$,
$\ev{K_1 K_2 \zetac^{2p+1}}$, and $\ev{K_2^2 \zetac^{2p}}$, with $p\ge0$.
Using the diagrammatic rules of \cref{eq:diagrules}, one gets
\begin{equation}
  \begin{aligned}
   &  \ev{K_1^2 \zetac^{2p+2}}  = \diagKoneKone \times W_{2p+2}
    \left(\diagZetaZeta\right)^{p+1} \\ & +
    (2p+1)\left(\diagKoneZeta\right) \times 2p
    \left(\diagKoneZeta\right) \\
    & \times W_{2p} \left(\diagZetaZeta\right)^p \\
    & = \dfrac{(2p+1)!}{p!\, 2^{p}}
    \ev{K_1^2}\ev{\zetac^2}^{p+1} + 4 \dfrac{(2p+2)!}{p!\,2^p}\ev{K}^2\ev{\zetac^2}^p,
  \end{aligned}
  \label{eq:m2K1zetap}
\end{equation}
together with
\begin{equation}
\begin{aligned}
  \ev{K_1K_2 \zetac^{2p+1}} & = \diagKtwoSelf \times (2p+1)
  \left(\diagKoneZeta\right) \\ & \times W_{2p}
  \left(\diagZetaZeta\right)^{p} \\
  & = -2 \dfrac{(2p+1)!}{p!\,2^{p}}
  \ev{K}^2 \ev{\zetac^2}^p,
\end{aligned}
\label{eq:K1K2zetap}
\end{equation}
and
\begin{equation}
\begin{aligned}
  \ev{K_2^2 \zetac^{2p}} & = \left[\left(\diagKtwoSelf\right)^2  + 2
  \left(\diagKtwoCycleII\right) \right] \\ & \times
  W_{2p}\left(\diagZetaZeta\right)^{p}  = \dfrac{5}{3} \dfrac{(2p)!}{p!\,2^p}
  \ev{K}^2 \ev{\zetac^2}^p.
\end{aligned}
\label{eq:m2K2zetap}
\end{equation}
Summing all the terms coming from the expansion of
\cref{eq:m2defOmegaK} gives the exact expression
\begin{equation}
\ev{\OmegaK^2} = \dfrac{1}{a^4 H^4} \left(\ev{K^2} + 80 \ev{K}^2\right)e^{8\ev{\zetac^2}}.
\label{eq:m2OmegaK}
\end{equation}
For a scale-invariant power spectrum, using
\cref{eq:evK,eq:varzetac,eq:m2K}, one obtains
\begin{equation}
  \ev{\OmegaKzero^2}  \simeq \dfrac{1}{9} \dfrac{\ksigma^4}{\azero^4 \Hzero^4}
  \calPstar\left(1+\dfrac{245}{12} \calPstar\right)e^{8\calPstar
    \Ninf}
   \simeq \dfrac{1}{9} \sigma^4 \calPstar.
\label{eq:m2OmegaKnum}
\end{equation}
Using \cref{eq:evOmegaKnum} for $\sigma=1$, the standard deviation of
$\OmegaKzero$ is given by
\begin{equation}
\sqrt{\ev{\OmegaKzero^2}-\ev{\OmegaKzero}^2} \simeq \dfrac{\sigma^2}{3}
\sqrt{\calPstar} \simeq 1.5 \times 10^{-5}.
\end{equation}

In summary, \cref{eq:evOmegaK,eq:m2OmegaK} show that, in a Universe
filled with cosmological fluctuations stretched over super-Hubble
scales, the curvature density parameter is not vanishingly small but
is promoted to a stochastic variable. At any time in the cosmic
history, we therefore expect an observer to measure a realization of
$\OmegaK$ dominated by its standard deviation, i.e., at about $1.5
\times 10^{-5}$. However, \cref{eq:K} makes explicit that $K$ is a
non-linear functional of $\zetac$. As such, even if $\zetac$ is of
Gaussian statistics, the probability distribution of $\OmegaK$ is,
\emph{a priori}, non-Gaussian. The rarity of extreme values of
$\OmegaK$ could, therefore, be affected by the higher moments, and we
now turn to their calculation.

\subsection{Higher moments}

All the higher moments $\ev{\OmegaK^n}$ with $n>2$ can be explicitly
calculated with the same method as the one employed for the mean value and
the variance. Expanding the exponential in series and using the binomial
expansion of $(K_1+K_2)^n$ shows that one has to determine the mean
value of combinations of the form $\ev{K_1^p K_2^q
  \zetac^m}=\ev{K_2^q}\ev{K_1^p \zetac^m}$. Those can all be
expressed in terms of powers of $\ev{\zetac^2}$, $\ev{K}$, and
$\ev{K^2}$ by using the diagrammatic rules of \cref{eq:diagrules}.

The only new subtlety consists in evaluating the terms in $\ev{K_2^q}$
that need to be decomposed into ``self-cycles''. For instance, the
third moment requires one to evaluate
\begin{equation}
\begin{aligned}
  \ev{K_2^3} & = 2^2 2!\left(\diagKtwoCycleIII\right) + 2 \left(\diagKtwoCycleII\right) \times 3
\left(\diagKtwoSelf\right) \\ & + \left(\diagKtwoSelf\right)^3,
\end{aligned}
\label{eq:meanK23}
\end{equation}
and one obtains
\begin{equation}
\begin{aligned}
  \ev{\OmegaK^3} & = -\dfrac{\ev{K}}{a^6 H^6}\left(39 \ev{K^2} +
  \dfrac{19430}{9} \ev{K}^2\right) e^{18 \ev{\zetac^2}}.
\end{aligned}
\end{equation}
Similarly, the fourth moment is given by
\begin{equation}
\begin{aligned}
\ev{\OmegaK^4} & = \dfrac{1}{a^8 H^8}\left(3 \ev{K^2}^2 + 1728
\ev{K^2}\ev{K}^2 \right. \\ & \left. + \dfrac{736682}{9}
\ev{K}^4\right) e^{32\ev{\zetac^2}},
\end{aligned}
\end{equation}
and so on and so forth. These expressions are not particularly
illuminating, but the leading-order terms of all the moments are
diagrammatically tractable, and one can show that, for a
scale-invariant power spectrum, the standardized moments $ \mutilde_n$ (the moments
divided by the $n^\textrm{th}$ power of the standard deviation) verify
\begin{equation}
\begin{aligned}
  \mutilde_{n=2p} &\simeq W_{n} \,e^{ \left( 2n^2-4n
    \right)\ev{\zetac^2}},\\ \mutilde_{n=2p+1} & \simeq n W_{n-1}
  \left(1+4n\right) \dfrac{\sqrt{\calPstar}}{2} \, e^{ \left( 2n^2-4n
    \right)\ev{\zetac^2}}.
\end{aligned}
\label{eq:mutilde}
\end{equation}
All odd standardized moments are suppressed by the factor
$\sqrt{\calPstar}$ with respect to the even ones. Moreover, provided
the exponential terms in \cref{eq:mutilde} are close to unity, i.e.,
for $n^2 \ev{\zetac^2} \ll 1$, the even moments exactly match the
ones associated with a Gaussian probability distribution. As such,
$\ev{\OmegaK^n}$ shows significant deviations compared to the
Gaussian expectations only for large values of $n^2 \gtrsim
1/\ev{\zetac^2}$. To better assess the effect of these higher moments, we next turn our attention to 
 the functional form of the $\OmegaK$'s probability
distribution.

\subsection{Probability distribution}

\begin{figure}
  \begin{center}
    \includegraphics[width=\figw]{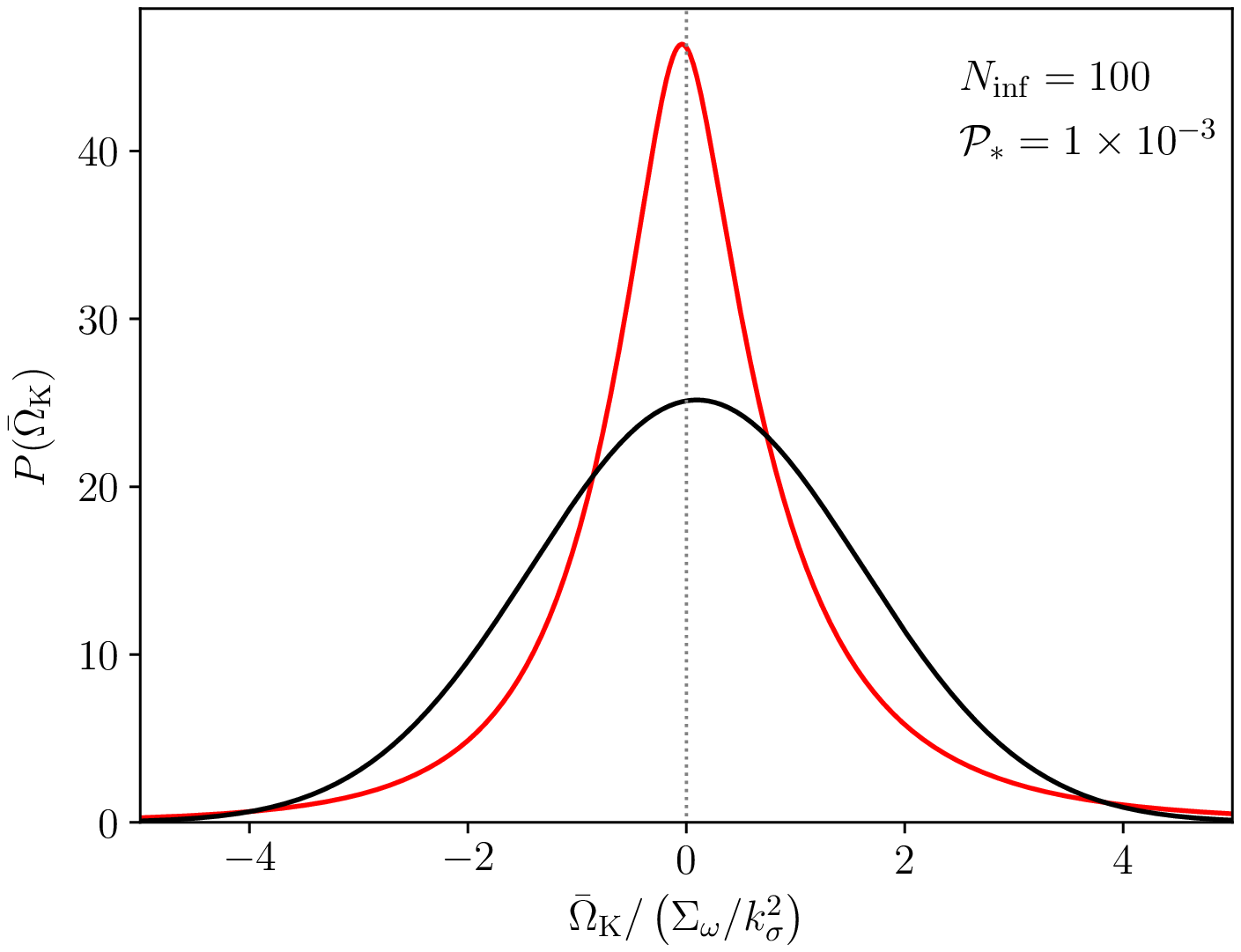}
    \caption{Probability distribution function for $\OmegaKbar =(a
      H/\ksigma)^2 \OmegaK$ (red curve) for unrealistically large
      values of $\calPstar = 10^{-3}$ (and $\Ninf=100$), compared to a
      Gaussian of same mean and variance (black curve). Notice that
      the most probable value of $\OmegaKbar$ is slightly negative
      whereas the mean value remains slightly positive.}
    \label{fig:proba}
  \end{center}
\end{figure}

The probability density function of $\OmegaK$ can be determined by
noticing that \cref{eq:K,eq:OmegaK} imply that $\OmegaK$ can be seen
as a non-linear functional over five
stochastic Gaussian variables, $\bXi \equiv (\zetac, \Delta \zetac,
\vnabla\zetac)$. As such, defining $\OmegaKbar \equiv (a^2 H^2/\ksigma^2)
\OmegaK$ and marginalizing over the five-dimensional space associated
with $\bXi$, one has
\begin{equation}
P(\OmegaKbar) = \int \dfrac{\dd[5]\bXi}{\left(2\pi\right)^{5/2}}
\, \delta\left(\OmegaKbar + \dfrac{K}{\ksigma^2} e^{-2 \zetac}\right)
\dfrac{e^{-\frac{1}{2} \bXi^{\uT} \bSigma^{-1} \bXi}}{\sqrt{\det \bSigma}},
\label{eq:Pmulti}
\end{equation}
where the five-dimensional covariance matrix $\bSigma$ is completely
determined by the diagrammatic rules of \cref{eq:diagrules}. All but
one integral appearing in \cref{eq:Pmulti} can be analytically reduced,
and, after some algebra, one obtains
\begin{equation}
\begin{aligned}
  P(\OmegaKbar) & = \dfrac{\ksigma^2}{4
    \pi}\sqrt{\dfrac{27\sqrt{2}\Sigmaw}{\ev{\zetac^2}\abs{\ev{K}}^{3}}} \int_{-\infty}^{+\infty} \dd x \,  e^{-\frac{x^2}{2\ev{\zetac^2}} + 2x}
   \\ & \times  e^{-\frac{1}{2}\omegakbar^2(\OmegaKbar,x)}\hermiteb{-\frac{3}{2}}{\dfrac{3
      \Sigmaw}{\sqrt{8} \abs{\ev{K}}} - \dfrac{\omegakbar(\OmegaKbar,x)}{\sqrt{2}}},
\end{aligned}
\label{eq:Pone}
\end{equation}
where we have defined
\begin{equation}
\begin{aligned}
  \Sigmaw^2 & \equiv \ev{K_1^2} - 4 \dfrac{\ev{K}^2}{\ev{\zetac^2}},\\
  \omegakbar(\OmegaKbar,x) &\equiv e^{2x}\dfrac{\OmegaKbar}{(\Sigmaw/\ksigma^2)} +
  \dfrac{2 \abs{\ev{K}}}{\ev{\zetac^2} \Sigmaw} x.
\end{aligned}
\end{equation}
In \cref{eq:Pone}, $\hermite{\nu}{x}$ stands for the generalized Hermite
polynomial of fractional order, defined from the parabolic cylinder
functions~\cite{1980tisp.book.G} as
$\hermite{\nu}{x}\equiv2^{\nu/2}e^{x^2/2} D_\nu(\sqrt{2}x)$. This
distribution shows that, for $\ev{\zetac^2} \simeq \calPstar \Ninf \ll
1$, one can use the approximation
\begin{equation}
e^{-\frac{x^2}{2\ev{\zetac^2}} + 2x} \simeq \sqrt{2\pi \ev{\zetac^2}}
e^{2\ev{\zetac^2}} \delta\left(x-2\ev{\zetac^2}\right),
\label{eq:deltax}
\end{equation}
to simplify the integral over $x$ in \cref{eq:Pone}. Remarking that,
in this limit, the argument of the Hermite function is dominated by
the first term, which is a constant scaling as $1/\sqrt{\calPstar}$,
$P(\OmegaKbar)$ is, therefore, close to a Gaussian distribution over the
quantity $\omegakbar(\OmegaKbar,2\ev{\zetac^2})$. In other words, for
$\ev{\zetac^2}\ll 1$, the distribution of $\OmegaKbar$ is almost
Gaussian, with a width given by $\Sigmaw/\ksigma^2
\simeq\sqrt{\calPstar}/3$ and a peak located at a very small
\emph{negative} value:
\begin{equation}
\left. \OmegaKbar \right|_{\max} \simeq \dfrac{4}{\ksigma^2} \ev{K}
e^{-4\ev{\zetac^2}} \simeq -\dfrac{2}{3} \calPstar.
\label{eq:OmegaKbar:max}
\end{equation}
For the curvature parameter today, one would get the most probable
value at $\left.\OmegaKzero\right|_{\max} \simeq -1.4\times 10^{-9}$,
a barely closed universe were we to interpret this number within a FLRW
metric with trivial topology. Let us notice the different sign than
the mean value of \cref{eq:evOmegaKnum}; the distribution is indeed
slightly skewed by the Hermite function. This can be seen in
\cref{fig:proba}, where we have plotted $P(\OmegaKbar)$ for an
unrealistically large value of $\calPstar=10^{-3}$. These distortions
are also apparent in the odd moments of \cref{eq:mutilde} which are,
as already noted, all proportional to $\sqrt{\calPstar}$.

\begin{figure}
  \begin{center}
    \includegraphics[width=\figw]{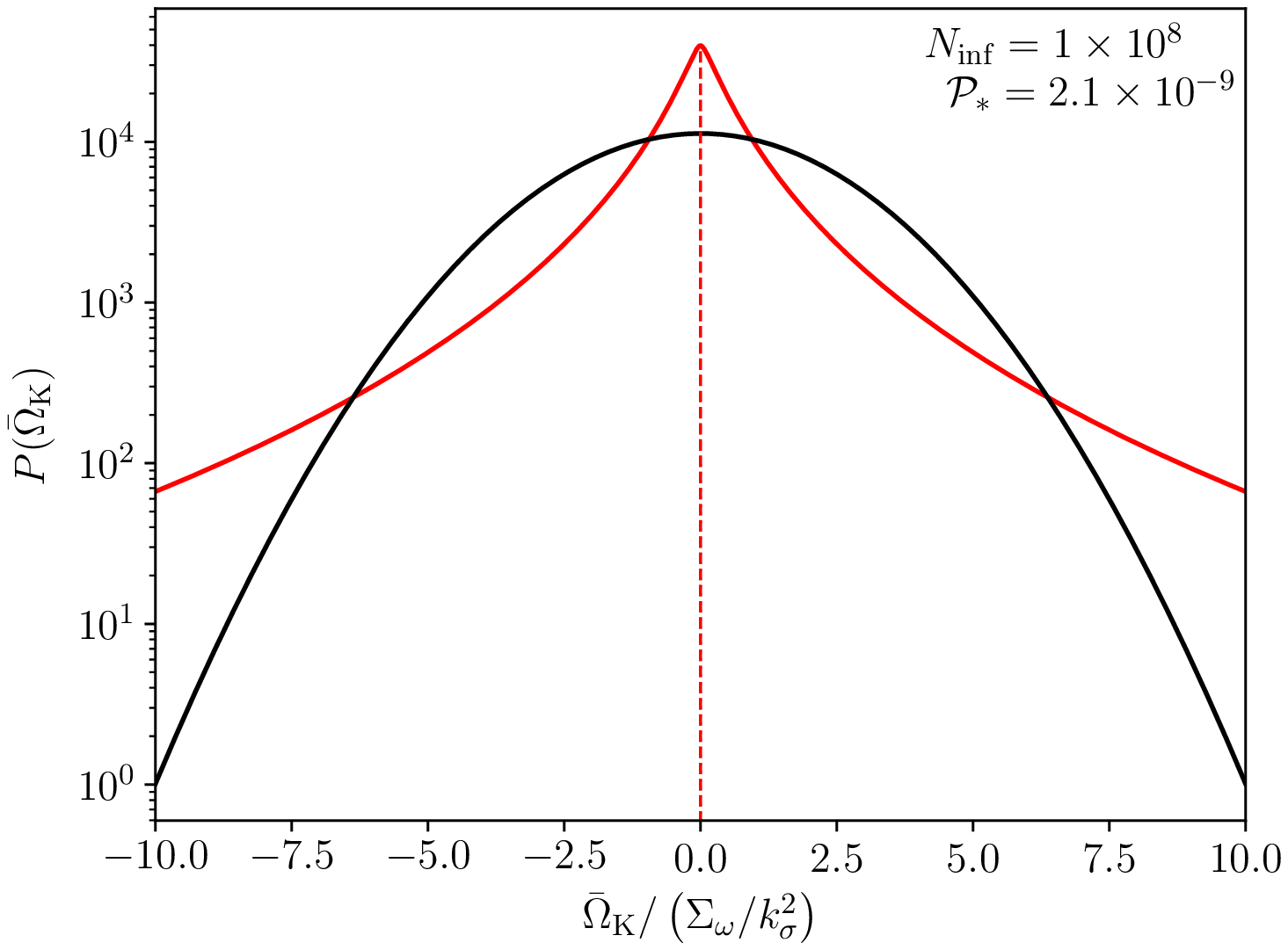}
    \caption{Probability distribution function for $\OmegaKbar =(a
      H/\ksigma)^2 \OmegaK$ (red curve) for the currently favored
      value of $\calPstar = 2.1 \times 10^{-9}$ and for a large number
      of {\efolds} $\Ninf=10^8$. The variance $\ev{\zetac^2}$ is no
      longer a small quantity and the distribution acquires heavy
      tails. Even though the width at half-maximum is
      $\order{\sqrt{\calPstar}}$, substantial values of $|\OmegaKbar|$
        are not rare anymore. For comparison, the black curve shows a
        Gaussian of same mean and variance.}
    \label{fig:tails}
  \end{center}
\end{figure}

When $\ev{\zetac^2} \simeq \calPstar \Ninf$ increases, \cref{eq:deltax} is
no longer accurate, and all the terms of \cref{eq:Pone} are
relevant. The distribution now acquires heavy tails, kicking in at
increasingly smaller values of $\left|\OmegaKbar\right|$ and erasing
the Gaussian profile in the neighborhood of
$\left.\OmegaKbar\right|_{\max}$. In \cref{fig:tails}, we have plotted
$P(\OmegaKbar)$, in logarithmic scales, for
$\calPstar=2.1\times10^{-9}$ and for a large number of {\efolds}
$\Ninf=10^8$.  These heavy tails imply that large values of $\vert
\OmegaKzero\vert$ are (much) more likely than what a Gaussian profile
would imply. Their existence is also manifest in the moments of
\cref{eq:mutilde} through the exponential coefficients involving
$\ev{\zetac^2}$. Such an effect is reminiscent of the non-linear
mapping of vacuum quantum fluctuations encountered in the context of
stochastic inflation~\cite{Pattison:2017mbe, Ezquiaga:2019ftu}.

Finally, let us mention that numerical computations of $\ev{\OmegaK}$
and $\ev{\OmegaK^2}$ based on using the distribution of \cref{eq:Pone}
do match the values we can get from \cref{eq:evOmegaK,eq:m2OmegaK}.

\section{Discussion}
\label{sec:discuss}

If inflation lasts for a long period, then substantial values of
$\OmegaKzero$ might be produced. Indeed, letting $\sigma^2\simeq
e^{-\ev{\zetac^2}}$ to implement the condition stated below
\cref{eq:ksigma}, \cref{eq:m2OmegaKnum} becomes
$\ev{\OmegaKzero^2}^{1/2} \simeq \sqrt{\calPstar}
e^{3\calPstar\Ninf}/3$. For this value not to exceed the current
observational bound $\vert\OmegaKzero\vert<3\times 10^{-3}$, with
$\calPstar=2.1\times 10^{-9}$ this leads to $\Ninf<7\times 10^8$. On
the one hand, this suggests that scenarios leading to phases of
inflation lasting for more than a billion {\efolds} might be
disfavored by current cosmological data. On the other hand, future
cosmological surveys, such as the ones using the neutral hydrogen line
at 21 cm, may possibly detect a non-vanishing curvature if inflation
actually lasted slightly less than a billion
{\efolds}~\cite{Witzemann:2017lhi}. Notice that the aforementioned
bound becomes more stringent if one accounts for the slightly red
observed spectral index.

Let us note, however, that when the above bound on $\Ninf$ is
saturated, $\ev{\zetac^2}\simeq 1.5$.  \emph{A priori}, our non-linear
formulas do not require $\ev{\zetac^2}$ to be small; hence, they can
still be used in that case. In particular, although one can see that
all the moments are becoming exponentially large with $\ev{\zetac^2}$,
\cref{eq:Pone} shows that $P(\OmegaKbar)$ remains
well defined. Nonetheless, the fact that the scale $\ksigma$ must be
set in a way that accommodates potentially large values of $\zeta$
suggests that our formalism may not be best suited in that case, and
the upper bound we have obtained on $\Ninf$ must be taken with
care. Moreover, for large $\ev{\zetac^2}$, possible backreaction
effects on super-Hubble scales could also induce deviations from
Gaussianity.

If inflation lasts even longer, $\OmegaK$ gets even larger, and our
formalism needs to be extended in at least two ways. First, when
$|\OmegaK|$ becomes of the order unity, or more, the metric associated
with \cref{eq:metric} is not acceptable anymore. For instance, a large
negative curvature density parameter would imply a compact manifold,
and this demands another coordinate system than the one in
\cref{eq:metric}. Second, when $|\OmegaK|$ becomes sizable, it opens
up a channel of backreaction of the curvature perturbation onto the
background dynamics, which, in turn, alters the inflationary
amplification of the curvature perturbations
themselves~\cite{Handley:2019anl, Letey:2022hdp}. This mechanism might
be tractable in an extended stochastic-inflation
formalism~\cite{Starobinsky:1986fx, Goncharov:1987ir,
  Starobinsky:1994bd, Vennin:2015hra, Grain:2020wro}, which we plan to
develop in a future work.
  
Finally, let us insist that our derivation of the statistics of
$\OmegaK$ is not rooted in any perturbative expansion of metric
coefficients. The assumptions made are that $\zeta$ is of Gaussian
statistics and conserved on super-Hubble scales. As such, our results
would be modified if curvature perturbations are non-Gaussian at
non-observably large scales. This is, strictly speaking, not excluded,
although it would require very specific early-universe models for
which curvature perturbations are Gaussian at observable scales today
(in order to satisfy the tight constraints on
non-Gaussianities~\cite{Planck:2019kim}) and non-Gaussian at larger
scales. Another hypothesis that could be broken is that $\zetac$ is
conserved by adiabaticity. The presence of entropic modes today could
invalidate this assumption, but, as for non-Gaussianities, their
presence during inflation is also disfavored by current data.

\begin{acknowledgments}
This work is supported by the ``Fonds de la Recherche Scientifique -
FNRS'' under Grant $\mathrm{N^{\circ}T}.0198.19$ as well as by the
Wallonia-Brussels Federation Grant ARC $\mathrm{N^{\circ}}19/24-103$.

\end{acknowledgments}

\bibliography{biblio}

\end{document}